%% file: main.tex
\newif\ifdoublecol

\doublecoltrue
\documentclass[conference]{IEEEtran}

\IEEEoverridecommandlockouts

\usepackage{cite}

\usepackage[cmex10]{amsmath}
\usepackage{amssymb}
\usepackage{amsmath}
\usepackage{enumerate}
\usepackage{multirow}
\usepackage{algpseudocode}
\usepackage[ruled,vlined]{algorithm2e}
\usepackage{amsthm}
\usepackage[caption=false,font=footnotesize]{subfig}
\usepackage{acro}
\usepackage{tikz}
\usepackage{xcolor}

\input{commands}

\input{abbrev}

\newtheorem{theorem}{Theorem}

\newcommand{\subparagraph}{}
\usepackage{titlesec}
\titlespacing\subsection{3pt}{4pt plus 4pt minus 2pt}{4pt plus 2pt minus 2pt}

\input{MohammadCommands}

\begin{document}

\title{
Low-complexity Multicast Beamforming for Multi-stream Multi-group Communications}

\author{\IEEEauthorblockN{Hamidreza Bakhshzad Mahmoodi, Bikshapathi Gouda, MohammadJavad Salehi and Antti T\"olli} \\
\IEEEauthorblockA{
    Centre for Wireless Communications, University of Oulu, 90570 Oulu, Finland \\
    \textrm{E-mail: \{firstname.lastname@oulu.fi\}}
    }
    	\thanks{
This work was supported by the Academy of Finland under grants no. 319059 (Coded Collaborative Caching for Wireless Energy Efficiency) and 318927 (6Genesis Flagship).}

}

\maketitle

\input{abstract}

\IEEEpeerreviewmaketitle

\input{main-txt}

\bibliographystyle{IEEEtran}
\bibliography{IEEEabrv,conf_short,jour_short,references}

\end{document}

%% file: commands.tex
\newtheorem{lemma}{Lemma}

\newcommand{\herm}{^{\mbox{\scriptsize H}}}

\newcommand{\vbar}{\raisebox{.17ex}{\rule{.04em}{1.35ex}}}
\newcommand{\vbarind}{\raisebox{.01ex}{\rule{.04em}{1.1ex}}}
\newcommand{\R}{\ifmmode{\rm I}\hspace{-.2em}{\rm R} \else ${\rm I}\hspace{-.2em}{\rm R}$ \fi}
\newcommand{\T}{\ifmmode{\rm I}\hspace{-.2em}{\rm T} \else ${\rm I}\hspace{-.2em}{\rm T}$ \fi}
\newcommand{\N}{\ifmmode{\rm I}\hspace{-.2em}{\rm N} \else \mbox{${\rm I}\hspace{-.2em}{\rm N}$} \fi}
\newcommand{\B}{\ifmmode{\rm I}\hspace{-.2em}{\rm B} \else \mbox{${\rm I}\hspace{-.2em}{\rm B}$} \fi}
\newcommand{\Hil}{\ifmmode{\rm I}\hspace{-.2em}{\rm H} \else \mbox{${\rm I}\hspace{-.2em}{\rm H}$} \fi}
\newcommand{\C}{\ifmmode\hspace{.2em}\vbar\hspace{-.31em}{\rm C} \else \mbox{$\hspace{.2em}\vbar\hspace{-.31em}{\rm C}$} \fi}
\newcommand{\Cind}{\ifmmode\hspace{.2em}\vbarind\hspace{-.25em}{\rm C} \else \mbox{$\hspace{.2em}\vbarind\hspace{-.25em}{\rm C}$} \fi}
\newcommand{\Q}{\ifmmode\hspace{.2em}\vbar\hspace{-.31em}{\rm Q} \else \mbox{$\hspace{.2em}\vbar\hspace{-.31em}{\rm Q}$} \fi}
\newcommand{\Z}{\ifmmode{\rm Z}\hspace{-.28em}{\rm Z} \else ${\rm Z}\hspace{-.28em}{\rm Z}$ \fi}



%% file: abbrev.tex
\DeclareAcronym{ADMM}{
    short = ADMM,
    long = alternating direction method of multipliers,
    list = Alternating Direction Method of Multipliers,
    class = abbrev
}

\DeclareAcronym{MGMC}{
    short = MGMC,
    long = multi-group multi-casting,
    list = multi-group multi-casting,
    class = abbrev
}

\DeclareAcronym{SGMC}{
    short = SGMC,
    long = single-group multi-casting,
    list = single-group multi-casting,
    class = abbrev
}
\DeclareAcronym{AoA}{
    short = AoA,
    long = angle-of-arrival,
    list = Angle-of-Arrival,
    class = abbrev
}

\DeclareAcronym{AoD}{
    short = AoD,
    long = angle-of-departure,
    list = Angle-of-Departure,
    class = abbrev
}

\DeclareAcronym{KKT}{
    short = KKT,
    long = Karush-Kuhn-Tucker,
    list = Karush-Kuhn-Tucker,
    class = abbrev
}

\DeclareAcronym{MMF}{
    short = MMF,
    long = max-min-fairness,
    list = max-min-fairness,
    class = abbrev
}

\DeclareAcronym{WMMF}{
    short = WMMF,
    long = weighted max-min-fairness,
    list = max-min-fairness,
    class = abbrev
}

\DeclareAcronym{BB}{
    short = BB,
    long = base band,
    list = Base Band,
    class = abbrev
}

\DeclareAcronym{BC}{
    short = BC,
    long = broadcast channel,
    list = Broadcast Channel,
    class = abbrev
}

\DeclareAcronym{BS}{
    short = BS,
    long = base station,
    list = Base Station,
    class = abbrev
}

\DeclareAcronym{BR}{
    short = BR,
    long = best response,
    list = Best Response, 
    class = abbrev
}

\DeclareAcronym{CB}{
    short = CB,
    long = coordinated beamforming,
    list = Coordinated Beamforming,
    class = abbrev
}

\DeclareAcronym{CC}{
    short = CC,
    long = coded caching,
    list = Coded Caching,
    class = abbrev
}

\DeclareAcronym{CE}{
    short = CE,
    long = channel estimation,
    list = Channel Estimation,
    class = abbrev
}

\DeclareAcronym{CoMP}{
    short = CoMP,
    long = coordinated multi-point transmission,
    list = Coordinated Multi-Point Transmission,
    class = abbrev
}

\DeclareAcronym{CRAN}{
    short = C-RAN,
    long = cloud radio access network,
    list = Cloud Radio Access Network,
    class = abbrev
}

\DeclareAcronym{CSE}{
    short = CSE,
    long = channel specific estimation,
    list = Channel Specific Estimation,
    class = abbrev
}

\DeclareAcronym{CSI}{
    short = CSI,
    long = channel state information,
    list = Channel State Information,
    class = abbrev
}

\DeclareAcronym{CSIT}{
    short = CSIT,
    long = channel state information at the transmitter,
    list = Channel State Information at the Transmitter,
    class = abbrev
}

\DeclareAcronym{CU}{
    short = CU,
    long = central unit,
    list = Central Unit,
    class = abbrev
}

\DeclareAcronym{D2D}{
    short = D2D,
    long = device-to-device,
    list = Device-to-Device,
    class = abbrev
}

\DeclareAcronym{DE-ADMM}{
    short = DE-ADMM,
    long = direct estimation with alternating direction method of multipliers,
    list = Direct Estimation with Alternating Direction Method of Multipliers,
    class = abbrev
}

\DeclareAcronym{DE-BR}{
    short = DE-BR,
    long = direct estimation with best response,
    list = Direct Estimation with Best Response,
    class = abbrev
}

\DeclareAcronym{DE-SG}{
    short = DE-SG,
    long = direct estimation with stochastic gradient,
    list = Direct Estimation with Stochastic Gradient,
    class = abbrev
}

\DeclareAcronym{DFT}{
	short = DFT,
	long = discrete fourier transform,
	list = Discrete Fourier Transform,
	class = abbrev
}

\DeclareAcronym{DoF}{
    short = DoF,
    long = degrees of freedom,
    list = Degrees of Freedom,
    class = abbrev
}

\DeclareAcronym{DL}{
    short = DL,
    long = downlink,
    list = Downlink,
    class = abbrev
}

\DeclareAcronym{GD}{
	short = GD, 
	long = gradient descent,
	list = Gradeitn Descent,
	class = abbrev
}

\DeclareAcronym{IBC}{
    short = IBC,
    long = interfering broadcast channel,
    list = Interfering Broadcast Channel,
    class = abbrev
}

\DeclareAcronym{i.i.d.}{
    short = i.i.d.,
    long = independent and identically distributed,
    list = Independent and Identically Distributed,
    class = abbrev
}

\DeclareAcronym{JP}{
    short = JP,
    long = joint processing,
    list = Joint Processing,
    class = abbrev
}

\DeclareAcronym{LOS}{
	short = LOS,
	long = line-of-sight,
	list = Line-of-Sight,
	class = abbrev
}

\DeclareAcronym{LS}{
    short = LS,
    long = least squares,
    list = Least Squares,
    class = abbrev
}

\DeclareAcronym{LTE}{
    short = LTE,
    long = Long Term Evolution,
    class = abbrev
}

\DeclareAcronym{LTE-A}{
    short = LTE-A,
    long = Long Term Evolution Advanced,
    class = abbrev
}

\DeclareAcronym{MIMO}{
    short = MIMO,
    long = multiple-input multiple-output,
    list = Multiple-Input Multiple-Output,
    class = abbrev
}

\DeclareAcronym{MISO}{
    short = MISO,
    long = multiple-input single-output,
    list = Multiple-Input Single-Output,
    class = abbrev
}

\DeclareAcronym{MSE}{
    short = MSE,
    long = mean-squared error,
    list = Mean-Squared Error,
    class = abbrev
}

\DeclareAcronym{MMSE}{
    short = MMSE,
    long = minimum mean-squared error,
    list = Minimum Mean-Squared Error,
    class = abbrev
}

\DeclareAcronym{mmWave}{
	short = mmWave,
	long = millimeter wave,
	list = Millimeter Wave,
	class = abbrev
}

\DeclareAcronym{MU-MIMO}{
    short = MU-MIMO,
    long = multi-user \ac{MIMO},
    list = Multi-User \ac{MIMO},
    class = abbrev
}

\DeclareAcronym{OFDM}{
    short = OFDM,
    long = orthogonal frequency division multiplexing,
    list = Orthogonal Frequency Division Multiplexing,
    class = abbrev
}

\DeclareAcronym{OTA}{
    short = OTA,
    long = over-the-air,
    list = Over-the-Air,
    class = abbrev
}

\DeclareAcronym{PSD}{
    short = PSD,
    long = positive semidefinite,
    list = Positive Semidefinite,
    class = abbrev
}

\DeclareAcronym{QoS}{
	short = QoS,
	long = quality of service,
	list = Quality of Service,
	class = abbrev
}

\DeclareAcronym{RCP}{
	short = RCP,
	long = remote central processor,
	list = Remote Central Processor,
	class = abbrev
}

\DeclareAcronym{RRH}{
    short = RRH,
    long = remote radio head,
    list = Remote Radio Head,
    class = abbrev
}

\DeclareAcronym{RSSI}{
    short = RSSI,
    long = received signal strength indicator,
    list = Received Signal Strength Indicator,
    class = abbrev
}

\DeclareAcronym{RX}{
	short = RX,
	long = receiver,
	list = Receiver,
	class = abbrev
}

\DeclareAcronym{SCA}{
    short = SCA,
    long = successive-convex-approximation,
    list = Successive-Convex-Approximation,
    class = abbrev
}

\DeclareAcronym{SG}{
    short = SG,
    long = stochastic gradient,
    list = Stochastic Gradient,
    class = abbrev
}

\DeclareAcronym{SNR}{
    short = SNR,
    long = signal-to-noise-ratio,
    list = Signal-to-Noise Ratio,
    class = abbrev
}

\DeclareAcronym{SDR}{
    short = SDR,
    long = semi-definite-relaxation,
    list = semi-definite-relaxation,
    class = abbrev
}

\DeclareAcronym{SINR}{
    short = SINR,
    long = signal-to-interference-plus-noise ratio,
    list = Signal-to-Interference-plus-Noise Ratio,
    class = abbrev
}

\DeclareAcronym{SOCP}{
	short = SOCP, 
	long = second order cone program,
	list = Second Order Cone Program,
	class = abbrev
}

\DeclareAcronym{SSE}{
    short = SSE,
    long = stream specific estimation,
    list = Stream Specific Estimation,
    class = abbrev
}

\DeclareAcronym{SVD}{
	short = SVD,
	long = singular value decomposition,
	list = Singular Value Decomposition,
	class = abbrev
}

\DeclareAcronym{TDD}{
	short = TDD,
	long = time division duplex,
	list = Time Division Duplex,
	class = abbrev
}

\DeclareAcronym{TX}{
	short = TX,
	long = transmitter,
	list = Transmitter,
	class = abbrev
}

\DeclareAcronym{UE}{
    short = UE,
    long = user equipment,
    list = User Equipment,
    class = abbrev
}

\DeclareAcronym{UL}{
    short = UL,
    long = uplink,
    list = Uplink,
    class = abbrev
}

\DeclareAcronym{ULA}{
	short = ULA,
	long = uniform linear array,
	list = Uniform Linear Array,
	class = abbrev
}

\DeclareAcronym{UPA}{
    short = UPA,
    long = uniform planar array,
    list = Uniform Planar Array,
    class = abbrev
}

\DeclareAcronym{WMMSE}{
    short = WMMSE,
    long = weighted minimum mean-squared error,
    list = Weighted Minimum Mean-Squared Error,
    class = abbrev
}

\DeclareAcronym{WMSEMin}{
    short = WMSEMin,
    long = weighted sum \ac{MSE} minimization,
    list = Weighted sum \ac{MSE} Minimization,
    class = abbrev
}

\DeclareAcronym{WBAN}{
	short = WBAN,
	long = wireless body area network,
	list = Wireless Body Area Network,
	class = abbrev
}

\DeclareAcronym{WSRMax}{
    short = WSRMax,
    long = weighted sum rate maximization,
    list = Weighted Sum Rate Maximization,
    class = abbrev
}

%% file: MohammadCommands.tex
\newcommand{\CK}[0]{{\mathcal{K}}}
\newcommand{\CL}[0]{{\mathcal{L}}}

\newcommand{\Bu}[0]{{\mathbf{u}}}

\newcommand{\Bw}[0]{{\mathbf{w}}}
\newcommand{\Bx}[0]{{\mathbf{x}}}
\newcommand{\By}[0]{{\mathbf{y}}}

\newcommand{\BK}[0]{{\mathbf{K}}}

\newcommand{\BW}[0]{{\mathbf{W}}}

\newcommand{\SfS}[0]{{\mathsf{S}}}



%% file: abstract.tex
\begin{abstract}

In this paper, assuming multi-antenna transmitter and receivers, we consider multicast beamformer design for the \ac{WMMF} problem in a multi-stream multi-group communication setup. Unlike the single-stream scenario, the \ac{WMMF} objective in this setup is not equivalent to maximizing the minimum weighted SINR due to the summation over the rates of multiple streams. Therefore, the non-convex problem at hand is first approximated with a convex one and then solved using \ac{KKT} conditions. Then, a practically appealing closed-form solution is derived, as a function of dual variables, for both transmit and receive beamformers. Finally, we use an iterative solution based on the sub-gradient method to solve for the mutually coupled and interdependent dual variables. The proposed solution does not rely on generic solvers and does not require any bisection loop for finding the achievable rate of various streams. As a result, it significantly outperforms the state-of-art in terms of computational cost and convergence speed.


\end{abstract}

\begin{IEEEkeywords}
Multi-stream multi-group communications; Multicast beamforming; Weighted max-min fairness
\end{IEEEkeywords}

%% file: main-txt.tex
\section{Introduction}
Wireless data communication is now playing a significant role in our everyday lives, and this importance will continue to grow with recent innovative applications such as autonomous driving and mobile immersive viewing. 
One of the key enablers for coping with the constantly growing mobile data traffic is the emergence of multi-antenna communications, which enables additional spatial \ac{DoF}, and hence, higher spectral efficiencies to be achieved~\cite{MIMO-magazine-2014}. Moreover, when the users' requests are correlated, one can benefit from multi-antenna multicasting techniques to serve multiple users within a group with the same content~\cite{SGMC-SDR-original-2006}. In multicasting, a single beamformer is used for a group of users, resulting in potentially higher bandwidth efficiency and transmission rate.
%
With the recent emergence of applications with correlating requests such as venue casting~\cite{Qualcomm} and mobile immersive viewing~\cite{mahmoodi2021non}, designing efficient multicasting techniques has also gained much attention from the research community.



In the basic multicasting setup, a single group of users is interested in the same common message. This basic setup is studied thoroughly in~\cite{SGMC-SDR-original-2006}, where beamformers are designed to minimize the total transmit power subject to a given \ac{SINR} at each user. It is shown that this problem, referred to as the \ac{QoS} problem, is NP-hard and can be solved efficiently using the \ac{SDR} method. 

An interesting extension to the basic multicast setup is proposed as the \ac{MGMC} problem, where various data streams are multicast to different groups of users simultaneously. Solving this problem through \ac{SDR} methods under different assumptions (e.g., total or per-user power constraint and centralized or decentralized settings) is extensively studied in the literature~\cite{MMF_QoS_MGMC_lefteris_2008,SDR_gossian-randomization_MGMC_TSP_2014,Coordinated-Multicast-beamforming-MC-2013,Tolli-Pennanen-Komulainen-TWC11}. However, \ac{SDR}-based approaches are computationally complex, and hence, alternative methods with reduced complexity have also gained interest. In~\cite{MGMC-SCA-ganesh-2017}, \ac{SCA} is used to design beamformers for both \ac{QoS} and \ac{MMF} problems in an \ac{OFDM} setup, and the solution is found iteratively using first-order Taylor expressions of \ac{SINR} terms. Similarly, in~\cite{MGMCmmf}, another solution based on \ac{KKT} conditions is proposed for \ac{QoS} and \ac{MMF} problems to make beamformer design complexity (almost) independent of the antenna count. From another perspective, other works in the literature have proposed semi-closed-form beamforming solutions to remove the dependency on generic solvers, hence potentially reducing the beamformer design complexity. For example, in~\cite{High-performance-adaptive-algorithm-SGMC-2015}, a descent direction method is used to solve the \ac{QoS} problem in a basic multicast setup, and in~\cite{consensus-ADMM-MC-2016,MGMC-ADMM-based-fast-algorithm-2017}, \ac{ADMM} is utilized to solve both \ac{QoS} and \ac{MMF} problems for the \ac{MGMC} setup. Similar works in this context are proposed also in~\cite{MGMC-large_antenna-meysam-sadeghi-2017,rate-spliting-MGMC-clarckx-2016,tcom-ahmet-common-message-MGMC-2019}.

In this paper, we address the \ac{MGMC} beamformer design problem in a \ac{MIMO} setup where both the transmitter and receivers have multiple antennas. 
In such a setup, multi-stream communication becomes possible, and the \ac{MMF} objective no longer corresponds to maximizing the minimum \ac{SINR} value as is generally considered in the literature. Instead, one has to maximize the sum of $\log(\mathrm{1+SINR})$ rate terms over multiple streams while guaranteeing the decodability of each stream at every user.
Due to these reasons, the state-of-art works such as~\cite{MGMCmmf, High-performance-adaptive-algorithm-SGMC-2015, consensus-ADMM-MC-2016, MGMC-ADMM-based-fast-algorithm-2017} are no longer suitable as they would require excessively long convergence times (due to inter-dependent bisection loops for calculating the rate of multiple streams). 
To address this issue, we first formulate the non-convex \ac{WMMF} problem for the considered system model, and then use \ac{SCA} to propose an approximate equivalent problem that is convex on either transmitter or receiver side but not jointly. 
For this convex problem, we use \ac{KKT} conditions to derive optimal beamformer expression in terms of dual variables. However, the dual variables are highly coupled and interdependent, preventing a closed-form solution. To solve this problem, we compute dual variables using a fast-converging iterative algorithm based on the sub-gradient method. Interestingly, the proposed algorithm outperforms other existing works even in the presence of single-antenna receivers, where the problem becomes equivalent to max-min \ac{SINR}. 
Simulation results show the superiority of the proposed algorithm over the state-of-the-art in terms of computation time and complexity.

Throughout the rest of the paper, we use bold-face lower- and upper-case letters to represent vectors and matrices, respectively. 
By $[K]$, we mean the set $\{1,2,...,K\}$. Other notations are defined as they are used throughout the text.

\section{System Model}\label{sec: system-model}
\subsection{Network Setup} \label{sec:sysmodel}
We consider a multi-group multicasting system where a single server with $N_{\mathrm T}$ transmitting antennas serves $K$ multi-antenna users. Every user $k \in [K]$ has $N_k$ receiving antennas. The user set is divided into $G$ non-overlapping \textit{groups}, such that every user appears in exactly one group and the users within the same group request the same multicast message. Let us use $\mathcal{K}_g$ to denote the set of user indices in multicast group $g \in [G]$. We use 
%
${\bf{W}}_g \in \mathbb{C}^{N_{\mathrm T}\times L_g}$ to denote the pre-coder matrix for users in $\mathcal{K}_g$, where $L_g$ represents the maximum number of independent streams that can be transmitted to these users ($L_g$ is a function of $N_{\mathrm T}, N_k$ and $K$).
The columns of $\BW_g$ are
stream-specific transmit beamformers for users in $\CK_g$, i.e., ${\bf{W}}_g = [{\bf{w}}_{g,1}, \dots, {\bf{w}}_{g,L_g}]$, where $\Bw_{g,l}$ is the beamforming vector for the $l$-th stream of group $g$. The channel matrix between user $k \in [K]$ and the transmitter is denoted by ${\bf{H}}_k \in \mathbb{C}^{N_k\times N_{\mathrm T}}$ and it is assumed to be perfectly known at the transmitter. Then, the received signal at user $k \in \mathcal{K}_g$ can be written as
\begin{equation} \label{eq:received signal_stream_specific}
    {\bf{y}}_k = {\bf{H}}_k {\bf{W}}_g {\bf{d}}_g+ \sum_{{\bar g} \neq g} {\bf{H}}_k {\bf{W}}_{\bar g} {\bf{d}}_{\bar g} + {\bf{z}}_k \; ,
\end{equation}
where ${\bf{d}}_g = [d_{g,1}, \dots, d_{g,L_g}]^T \in \mathbb{C}^{L_g}$ is the transmitted data vector to multicast group $g \in [G]$ with $\mathbb{E}\{{\bf{d}}_g {\bf{d}}\herm _g\} = {\bf{I}}_{L_g}$, and ${\bf{z}}_k \sim \mathbb{CN}({\bf{0}}, \sigma_k^2{\bf{I}}_{N_k})$ is the additive white Gaussian noise at user $k$ with noise variance $\sigma_k^2$. The estimate of $d_{g,l}$ at user $k$ is given by $\hat{d}_{k,l} = {\bf{u}}\herm _{k,l} {\bf{y}}_k$, where ${\bf{u}}_{k,l} \in \mathbb{C}^{N_k}$ denotes the corresponding linear receive beamforming vector. The \ac{MSE} for stream $l$ of user $k \in \mathcal{K}_g$ can then be written as
\begin{align}
\label{eq:MSE}
        \epsilon_{k,l} ({\bf{W}},{\bf{u}}_{k,l}) = & |1-{\bf{u}}\herm _{k,l} {\bf{H}}_k {\bf{w}}_{g,l}|^2 + \\
        &\sum_{\bar g \in [G]} \sum_{\substack{j \in [L_{\bar g}] \\ (\bar g,j) \neq (g,l)}} |{\bf{u}}\herm _{k,l} {\bf{H}}_k {\bf{w}}_{\bar g,j}|^2 \nonumber + \sigma^2 \| {\bf{u}}_{k,l}\|^2 \; ,
\end{align}
where ${\bf{W}} := [{\bf{W}}_1, \dots, {\bf{W}}_G]$. Note that \eqref{eq:MSE} is convex on $\{{\bf{u}}_{k,l}\}$ or  $\{{\bf{w}}_{g,l}\}$ but not on both at the same time. Finally, the \ac{SINR} for stream $l$ at user $k \in \mathcal{K}_g$ can be written as
\begin{equation}\label{eq:SINR}
    \gamma_{k,l} \! = \frac{|{\bf{u}}\herm _{k,l} {\bf{H}}_k {\bf{w}}_{g,l}|^2}{\sum_{\bar g \in [G]}\sum_{\substack{j \in [L_{\bar g}] \\ (\bar g,j) \neq (g,l)}} |{\bf{u}}\herm _{k,l} {\bf{H}}_k {\bf{w}}_{\bar g,j}|^2 +  \sigma^2 \| {\bf{u}}_{k,l}\|^2} \, .
\end{equation}

\subsection{Problem Formulation}\label{sec:problemformulation_streamspecific}
First, we consider linear beamformers to simplify the beamforming process. Then, in Section~\ref{sec:problemformulation_upperbound}, we remove this assumption and consider a more general model. The performance gap of these two models is compared in Section~\ref{sec:simulation}.

The objective is to achieve the weighted fairness among different groups. For every group $g \in [G]$, we aim to maximize the sum-rate over all its $L_g$ streams. Moreover, as
each data stream $d_{g,l}$ is requested by all users in group $g$, its corresponding transmission rate should be assigned such that every user in $\CK_g$ can decode it. Hence, we need to solve 
\begin{subequations}\label{eq:main-MMF}
\begin{align}
\SfS_{0} \ : \quad &\max_{{\bf{w}}_{g,l}, {\bf{u}}_{k,l}} \min_{g \in [G]}\alpha_{g}\sum_{l \in [L_g]} \min_{k \in \mathcal{K}_g} \log(1+\gamma_{k,l}) \\
    \textrm{s.t.} \quad &\sum_{g \in [G]}\sum_{l\in [L_g]}\|{\bf{w}}_{g,l}\|^2 \leq P_{T}\label{const:power},
\end{align}
\end{subequations}
where $\alpha_g$ is the associated weight for group $g$ and $P_T$ is the total available power at the transmitter. Problem $\SfS_{0}$ is non-convex on both receive and transmit beamformers, but following similar steps as in \cite{jarkojornal2016}, can be solved for $\Bu_{k,l}$ (while fixing $\Bw_{g,l}$). The result is the standard linear \ac{MMSE} receiver
\begin{equation}\label{eq:MMSE_receiver}
    {\bf{u}}_{k,l} = \left({\bf{H}}_k{\bf{W}}{\bf{W}}\herm {\bf{H}}\herm _k+ \sigma_k^2{\bf{I}}\right)^{-1}{\bf{H}}_k{\bf{w}}_{g,l} \; .
\end{equation} 
Using~\eqref{eq:MMSE_receiver} in~\eqref{eq:MSE}, the SINR terms in~\eqref{eq:SINR} can be written as $\gamma_{k,l} = \epsilon^{-1}_{k,l} -1, \ \forall(k,l)$. Thus, writing the rate expression in~\eqref{eq:main-MMF} in terms of \ac{MSE} and relaxing the objective, the problem $\SfS_{0}$ can be reformulated as
\begin{subequations}\label{eq:MMF_step2}
\begin{align}
\SfS_{1} \ : \quad &\max_{{\bf{w}}_{g,l}, r_{g,l}} r_c\\
    \textrm{s.t.} \quad &r_c \leq \alpha_g \sum_{l \in [L_g]} r_{g,l} \, , \quad \forall g \in [G] \; , \label{const: sum_rate}\\
    &r_{g,l} \leq \log(\epsilon^{-1}_{k,l}) \, ,  \quad \forall (g,k \in \mathcal{K}_g, \, l \in [L_g]) \; , \label{const:MSE terms}\\
    & \text{and the power constraint in~\eqref{const:power}} \; . \nonumber
\end{align}
\end{subequations}
Note that \ac{MSE} constraint~\eqref{const:MSE terms} in $\SfS_{1}$ is still non-convex. To relax this constraint, we use auxiliary variables $t_{k,l}$ satisfying 
\begin{equation}\label{eq:MSE_aprox}
    \epsilon_{k,l} \leq [f(t_{k,l})]^{-1},
\end{equation}
where $f(t_{k,l})$ is a monotonic and continuously differentiable function that is Lipschitz continuous (hence, has finite first-order approximation coefficients), log-concave on its domain (i.e., $t \in \{x | f(x) \in [1, \infty]\}$), and equipped with convex multiplicative inverse (i.e., $[f(t)]^{-1}$ is convex on $t \in \{x| f(x) \in [1,\infty]\}$).
%
The domain of $f(\cdot)$ is dictated by the range of \ac{MSE} values, i.e., $\epsilon_{k,l} \in (0, 1] \, ,\forall (k,l)$. There are different classes of functions satisfying these conditions (c.f.~\cite{jarkojornal2016}). For convenience, in this paper we assume $f(t_{k,l}) = 2^{t_{k,l}}$.
Applying~\eqref{eq:MSE_aprox} into~\eqref{const:MSE terms} the problem $\SfS_1$ can be written as
\begin{subequations}\label{eq:MMF_relaxmmse}
\begin{align}
\SfS_{2} \ : \quad &\max_{\substack{{\bf{w}}_{g,l}, r_{g,l}, t_{k,l}}} r_c \\
    \textrm{s.t.} \quad  &r_{g,l} \leq \log(f(t_{k,l})), \quad \forall (g,k \in \mathcal{K}_g, \, l) \; , \label{const: per-stream rate}\\
    &\epsilon_{k,l} \leq [f(t_{k,l})]^{-1} \label{const:MSE terms_relax} \; ,\\
    &   \text{and the constraints in~\eqref{const:power} and~\eqref{const: sum_rate}}  \; . \nonumber
\end{align}
\end{subequations}
The constraint~\eqref{const:MSE terms_relax} is still non-convex. However, as $[f(t_{k,l})]^{-1}$ is convex, it can be lower bounded by its first-order Taylor approximation. Thus, $\SfS_{2}$ can be further approximated by
\begin{subequations}\label{eq:MMF_convex}
\begin{align}
\SfS_{3} \ : \quad &\max_{\substack{{\bf{w}}_{g,l}, r_{g,l}, t_{k,l}}} r_c \\
    \textrm{s.t.} \quad &\epsilon_{k,l} \leq \overline{a}_{k,l}t_{k,l}+\overline{b}_{k,l}, \quad \forall (k \in \mathcal{K}_g, \, l) \; , \label{const:MSE terms_relaxfx}\\
    &\text{and the constraints in~\eqref{const:power}, \eqref{const: sum_rate}, and~\eqref{const: per-stream rate}} \; , \nonumber
\end{align}
\end{subequations}
where $[f(t_{k,l})]^{-1}$ in~\eqref{const:MSE terms_relax} is replaced by its first-order Taylor approximation at $\overline{t}_{k,l}$ (the corresponding $t_{k,l}$ value in the previous iteration), and 
\begin{equation}
\label{eq:a_k-b_k}
    \overline{a}_{k,l} = -\frac{f^{'}(\overline{t}_{k,l})}{[f(\overline{t}_{k,l})]^{2}}, \;  \overline{b}_{k,l} = \frac{f(\overline{t}_{k,l}) + \overline{t}_{k,l}f^{'}(\overline{t}_{k,l})}{[f(\overline{t}_{k,l})]^{2}} \; .
\end{equation}
The problem $\SfS_{3}$ is convex for transmit beamformers $\Bw_{g,l}$ and can be solved optimally.

\section{Algorithmic Solutions}\label{sec:solutions}

\subsection{CVX-based Solution} \label{solution: CVX_streamspecific}
Since problem $\SfS_3$ in~\eqref{eq:MMF_convex} is convex (on either transmit or receive beamformers but not jointly), it can be directly handled by generic solvers such as CVX. In this paper, we consider the CVX solution as our benchmark. The general procedure for solving the problem with CVX is provided in Algorithm~\ref{algorithm:MMF_cvx}.
In a nutshell, we first assign a random initial value for every ${\bf{w}}^{0}_{g,l}$, such that the power constraint in~\eqref{const:power} is met. Next, we compute receive beamformers ${\bf{u}}_{k,l}$ 
using~\eqref{eq:MMSE_receiver}, \ac{MSE} values ${\overline{\epsilon}_{k,l}}$ using~\eqref{eq:MSE}, and $\overline{{t}}_{k,l}$ values using $\overline{t}_{k,l} = f^{-1}({\overline{\epsilon}^{-1}_{k,l}})$. Then, assuming the receive beamformers are fixed, we use CVX to solve~\eqref{eq:MMF_convex} and update transmit beamformers ${\bf{w}}_{g,l}$, which are later used to update receive beamformers and MSE values for the next iteration. The whole process is repeated until the convergence is achieved. Following a similar argument as in~\cite{MGMC-ADMM-based-fast-algorithm-2017}, the computation complexity of CVX per iteration is approximately equal to $\mathcal{O}\big((N_{\mathrm T} + \sum_g|\mathcal{K}_g|L_g)^{3.5}\big)$.

\begin{algorithm}
\SetAlgoLined
\KwResult{${\bf{w}}_{g,l}, {\bf{u}}_{k,l}$, $r_c$, $r_{g,l}$, $t_{k,l}$}
Choose random vectors for ${\bf{w}}_{g,l}$ such that~\eqref{const:power} is met\; 
 \While{convergence is not met}{
 Compute ${\bf{u}}_{k,l}$ from~\eqref{eq:MMSE_receiver}, using ${\bf{w}}_{g,l}$;
 
 Compute $\overline{\epsilon}_{k,l}$ from~\eqref{eq:MSE}, using ${\bf{w}}_{g,l}$, ${\bf{u}}_{k,l}$;
 
 Compute $\overline{t}_{k,l} = f^{-1}({\overline{\epsilon}^{-1}_{k,l}})$;
 
 Compute $\overline{a}_{k,l}$ and $\overline{b}_{k,l}$ from~\eqref{eq:a_k-b_k}, using $\overline{t}_{k,l}$; 
 
 Solve~\eqref{eq:MMF_convex} with CVX and find ${\bf{w}}_{g,l}, r_c, r_{g,l}, t_{k,l}$.
 }
 \caption{CVX-based solution for \ac{WMMF}}
 \label{algorithm:MMF_cvx}
\end{algorithm}
\subsection{The Proposed Iterative Solution}\label{sec:iterative_solution}
We propose a low-complexity, \ac{KKT}-based solution for computing transmit and receive beamformers iteratively. We first need the Lagrangian function for problem $\SfS_3$ in~\eqref{eq:MMF_convex}, for which we have
%
\begin{align} \label{eq:mmf-lagrangian}
    \mathcal{L}(\cdot) = &-r_c   + \sum_{g \in [G]}\zeta_g (r_c-\alpha_g \sum_{l \in [L_g]} r_{g,l} ) \nonumber \\
    & + \mu  \big( \sum_{g \in [G]}\sum_{l\in[L_g]} \|{\bf{w}}_{g,l}\|^2 - P_{T} \big) \nonumber\\ 
    & + \sum_{g \in [G]}\sum_{k \in \mathcal{K}_g} \sum_{l \in [L_g]} v_{k,l} \big(r_{g,l} - \log(f(t_{k,l})) \big) \nonumber\\
    & +\sum_{g \in [G]}\sum_{k \in \mathcal{K}_g} \sum_{l \in [L_g]} \lambda_{k,l}\left(\epsilon_{k,l} - \overline{a}_{k,l}t_{k,l}-\overline{b}_{k,l}\right) \; ,
\end{align}
where $\CL(\cdot) := \mathcal{L}(\lambda_{k,l},v_{k,l} ,\mu, \zeta_g, t_{k,l},{\bf{W}},r_c,r_{g,l})$.
The dual variables $\mu$, $\zeta_g$, $v_{k,l}$, and $\lambda_{k,l}$ are related to the power, common rate, stream-specific rate, and \ac{MSE} constraints respectively.

\begin{theorem}
With fixed receive beamformers $\overline{{\bf{u}}}_{k,l}$ and the auxiliary function $f(t_{k,l}) = 2^{t_{k,l}}$, the following primal and dual variables satisfy the \ac{KKT} conditions at the optimal point
\begin{subequations}
\begin{align}
        &{\bf{w}}^{*}_{g,l}(\lambda_{k,l}^{*},\mu^{*},\bar{{\bf{U}}}) = \nonumber \\
        & \qquad \left({\bf{H}}\herm \bar{{\bf{U}}} \bar{{\bf{U}}}\herm {\bf{H}} + \mu^{*}{\bf{I}}\right)^{-1}\sum_{k \in \mathcal{K}_g}\lambda^{*}_{k,l}{\bf{H}}_{k}\herm \overline{{\bf{u}}}_{k,l} \; ;  \label{eq:opt_transmit}\\
        &r^{*}_{g,l}(v_{k,l}^{*},\epsilon^{*}_{k,l}) = \frac{\sum_{k \in \mathcal{K}_g} v^{*}_{k,l}\log(\frac{1}{\epsilon^{*}_{k,l}})}{\sum_{k \in \mathcal{K}_g}v^{*}_{k,l}}, \label{eq:opt_rate} \quad \forall (g, l) \; ; \\
        &r^{*}_c(v_{k,l}^{*},\epsilon^{*}_{k,l}) = \sum_{g \in [G]}\sum_{k \in \mathcal{K}_g}\sum_{l \in [L_g]} v^{*}_{k,l}\log(\frac{1}{\epsilon^{*}_{k,l}}) \; ; \label{eq:opt_comrate}\\
        &\zeta^{*}_g = {\alpha_g}^{-1} \sum_{k \in \mathcal{K}_g}v^{*}_{k,l}, \quad \forall g \; ; \label{eq:opt_zeta}\\
        &\sum_{g \in [G]} \alpha^{-1}_{g} \sum_{k \in \mathcal{K}_g} v^{*}_{k,l} = 1, \quad \forall l \; ; \label{eq:opt_v}\\
        &\lambda^{*}_{k,l} = \frac{v^{*}_{k,l}}{\epsilon^{*}_{k,l}}, \quad  \forall k \in \CK_g, \forall (g,l) \; ; \label{eq:opt_lambda}\\
        & \mu^{*}(\lambda^{*}, \overline{{\bf{u}}}_{k,l}) = \frac{1}{P_T}\sum_{g \in [G]}\sum_{k \in \mathcal{K}_g} \sum_{l\in [L_g]} \lambda^{*}_{k,l}\overline{{\bf{u}}}_{k,l}\herm \overline{{\bf{u}}}_{k,l} \; ; \label{eq:opt_mu}
\end{align}
\end{subequations}
where $\epsilon^{*}_{k,l} = \epsilon_{k,l}({\bf{W}}^{*},\overline{\Bu}_{k,l})$, $\mu^{*} = \mu(\lambda^{*}_{k,l}, \overline{{\bf{u}}}_{k,l})$, ${\bf{H}}:=[{\bf{H}}_{1}, \dots, {\bf{H}}_{K}]$, $\overline{\Bu}_{k,l}$ is calculated using~\eqref{eq:MMSE_receiver}, and $\bar{{\bf{U}}}$ is a block-diagonal matrix with elements $\bar{{\bf{U}}}_k$,  where $k \in [K]$ and $\bar{{\bf{U}}}_k = [\sqrt{\lambda_{k,1}}\overline{{\bf{u}}}_{k,1}, \dots, \sqrt{\lambda_{k,L_g}}\overline{{\bf{u}}}_{k,L_g}]$. Note that~\eqref{eq:opt_zeta} and~\eqref{eq:opt_v} each represent a set of $\sum_{g \in [G]}L_g$ conditions. 

\begin{proof}
The condition~\eqref{eq:opt_transmit} on optimal transmit beamformers 
results from the stationary KKT condition with respect to 
${\Bw}_{g,l}$, i.e., $\nabla_{{\bf{w}}_{g,l}}\mathcal{L}(.)|_{{\bf{w}}_{g,l} \to {\bf{w}}^{*}_{g,l}} = 0$. Similarly, conditions~\eqref{eq:opt_zeta},~\eqref{eq:opt_v}, and~\eqref{eq:opt_lambda} result from stationary KKT conditions with respect to $r_{g,l}$, $r_c$, and $t_{k,l}$, respectively (note that to achieve~\eqref{eq:opt_v}, we have to once use~\eqref{eq:opt_zeta} to replace $\zeta^{*}_g$). To derive optimal stream-specific rates in~\eqref{eq:opt_rate}, we first update $t_{k,l}$ using $t^{*}_{k,l} = f^{-1}(\epsilon^{-1}_{k,l})$. Then, we use complementary slackness on~\eqref{const: per-stream rate} and sum over all the users within group $g$, to get $\sum_{k \in \mathcal{K}_g}v^{*}_{k,l} ( r_{g,l} -\log(f(t^{*}_{k,l})t)) = 0$. This results in
\begin{equation}\label{eq:opt_rgl_direct}
    r_{g,l} = \frac{-\sum_{k \in \mathcal{K}_g} v^{*}_{k,l}\log(\epsilon^{*}_{k,l})}{\alpha_g\zeta^{*}_g} \; ,
\end{equation}
which can then yield~\eqref{eq:opt_rate} by simply replacing $\zeta^{*}_g$ with~\eqref{eq:opt_zeta}.
Similarly, we can use complementary slackness on~\eqref{const: sum_rate} and sum over all the groups, i.e., $\sum_{g}\zeta^{*}_g (r_c-\alpha_g \sum_{l = 1}^{L}r_{g,l} ) = 0$, and then replace $r_{g,l}$ from~\eqref{eq:opt_rgl_direct} to get~\eqref{eq:opt_comrate}.
Finally, the dual variable $\mu^{*}$ is derived using similar steps as in~\cite{tcom-ahmet-common-message-MGMC-2019}. To save space, the steps are not repeated here and are left for the extended version of this paper. 
\end{proof}
\end{theorem}

From ~\eqref{eq:opt_transmit}-\eqref{eq:opt_mu}, we can see that
there exist closed-form solutions for all variables except for $v_{k,l}$. From~\eqref{eq:opt_v}, it can be seen that variables $v_{k,l}$ are interdependent, and hence, proposing a closed-form solution is infeasible. However, we can still use sub-gradient method to update $v_{k,l}$ (c.f.~\cite{jarkojornal2016,dileepjournal2020}). The following Lemma clarifies this procedure.
\begin{lemma}\label{lem:sub_gradian}
The gradient of $\CL(\cdot)$ in~\eqref{eq:mmf-lagrangian} with respect to $v_{k,l}$ at the point given by $\overline{r}_c$, $\overline{r}_{g,l}$, and $\epsilon_{k,l}$ can be written as
\begin{equation}\label{eq:sub_gradient_gradient}
    \nabla_{v_{k,l}}\mathcal{L}(\cdot) = \frac{\overline{r}_c - \alpha_g\sum_{l \in [L_g]}\overline{r}_{g,l}}{\alpha_g L_g} + \overline{r}_{g,l} + \log\left(\epsilon_{k,l}\right).
\end{equation}
\begin{proof}
Since~\eqref{eq:opt_zeta} is true for any $l \in [L_g]$, we can sum its both sides over all $L_g$ streams to get
\begin{equation}\label{eq:opt_zeta_sum}
    \zeta_g = \frac{\sum_{k \in \mathcal{K}_g}\sum_{l \in [L_g]}v_{k,l}}{\alpha_gL_g} \; .
\end{equation}
Now, we can replace $\zeta_g$ in~\eqref{eq:mmf-lagrangian} with its equivalent in~\eqref{eq:opt_zeta_sum}, and take the derivative with respect to $v_{k,l}$ to get~\eqref{eq:sub_gradient_gradient}. 
\end{proof}
\end{lemma}
Using Lemma~\ref{lem:sub_gradian}, the sub-gradient update for dual variables
$v_{k,l}$ can be done using
\begin{equation}\label{eq:sub_gradient_v}
    v^{(n)}_{k,l} = \left[v^{(n-1)}_{k,l} + \beta\nabla_{v_{k,l}}\mathcal{L}(\overline{r}_c, \overline{r}_{g,l},\epsilon_{k,l})\right]^+, \ \ \forall(k,g,l)
\end{equation}
where $\beta > 0$ is the step size and $[x]^+ := \max(x,0)$. 

In Algorithm~\ref{algorithm:MMF}, we have outlined the general procedure of the proposed iterative solution. 
As an additional explanation, we first choose a set of random transmit beamformers ${\bf{w}}_{g,l}$ such that the power constraint in~\eqref{const:power} is met. 
Also, assuming zero common rate for each stream (i.e., $\epsilon_{k,l} = 1$, $\forall (k,l)$), we initialize rate and \ac{MSE} dual variables ($\lambda_{k,l}, v_{k,l}$) with $\frac{\alpha_g}{K}$ (so that~\eqref{eq:opt_zeta}-\eqref{eq:opt_lambda} are satisfied).
Then, we iteratively compute primal and dual variables using~\eqref{eq:MMSE_receiver}, \eqref{eq:opt_transmit}-\eqref{eq:sub_gradient_v}, until the convergence is met. In Algorithm~\ref{algorithm:MMF}, the maximum value for inner and outer loop iterations for SCA and sub-gradient updates are denoted by $I_{in}$ and $I_{out}$, respectively. It is worth noting that since~\eqref{eq:opt_mu} is valid only for optimal $\lambda^{*}_{k,l}$ values, it may not satisfy the power constraint in~\eqref{const:power} at every iteration. Thus, in each iteration, we have to use the bisection method to compute a $\mu$ value satisfying~\eqref{const:power}. 
Moreover, 
from the complexity perspective, the dominant term in the proposed iterative solution is the inversion of the $N_{\mathrm T} \times N_{\mathrm T}$ matrix in~\eqref{eq:opt_transmit}, which requires the complexity of $\mathcal{O}((N_{\mathrm T} + \sum_{g \in [G]}|\mathcal{K}_g|L_g)N^{2}_{\mathrm T})$. As a result, the complexity of the proposed method scales linearly with the number of users in the network (or the number of streams), making it suitable for large networks.

 \begin{algorithm}
\SetAlgoLined
\KwResult{${\bf{w}}_{g,l}, {\bf{u}}_{k,l}, \lambda_{k,l}$, $r_{g,l}$, $r_c$, $\mu$, $v_{k,l}$, $\epsilon_{k,l}$}

 Set $\hat{i} \gets 0$, $v_{k,l}^{(0)} \gets \frac{\alpha_g}{K}, \lambda_{k,l}^{(0)} \gets \frac{\alpha_g}{K}, \  \forall (l,k)$;
 
 Choose random vectors for ${\bf{w}}_{g,l}$ such that~\eqref{const:power} is met; 
 
 \While{convergence not met $\mathrm{and}$ $\hat{i} < $ {$I_{out}$}}{
 Set $\hat{i} \gets \hat{i} + 1$, $i \gets 0$ ;
 
 Compute ${\bf{u}}_{k,l}$ from~\eqref{eq:MMSE_receiver}, using ${\bf{w}}_{g,l}$;
 
 \While{convergence not met $\mathrm{and}$ $i < $ {$I_{in}$}}{
   Set $i \gets i+1$;
   
 Solve ${\bf{w}}_{g,l}$ from~\eqref{eq:opt_transmit}, using ${\bf{u}}_{k,l},\lambda^{(i-1)}_{k,l}, \mu$;
 
 $\quad \triangleright$ $\mu$ is found using bisection, meeting~\eqref{const:power}.
 

  Compute $\epsilon_{k,l}$ from~\eqref{eq:MSE}, using ${\bf{w}}_{g,l}$, ${\bf{u}}_{k,l}$;

  Compute $r_{g,l}$ from~\eqref{eq:opt_rate}, using $\epsilon_{k,l}, v_{k,l}^{(i-1)}$;
  
  Compute $r_c$ from~\eqref{eq:opt_comrate}, using $\epsilon_{k,l}, v_{k,l}^{(i-1)}$;
  
  Update $v_{k,l}^{(i)}$ from~\eqref{eq:sub_gradient_v}, using  $r_c, v_{k,l}^{(i-1)}\!, r_{g,l}, \epsilon_{k,l}$;
  
  Normalize $v_{k,l}^{(i)}$ by $\sum_{g \in [G]} \alpha^{-1}_g\sum_{k \in \mathcal{K}_g} v_{k,l}^{(i)}$;
  
  $\quad \triangleright$ This is done to satisfy~\eqref{eq:opt_v}.
  
  Update $\lambda_{k,l}^{(i)}$ from~\eqref{eq:opt_lambda}, using  $v_{k,l}^{(i)}, \epsilon_{k,l}$;
}


 }
 \caption{Iterative algorithm for \ac{WMMF}}
 \label{algorithm:MMF}
\end{algorithm}

\subsection{Upper-bound}\label{sec:problemformulation_upperbound}
In section~\ref{sec: system-model}, we modeled the transmitted signal for group $g$ as ${\bf{x}}_{g} = {\bf{W}}_g {\bf{d}}_g$. As a result, the maximum rank of the transmit covariance matrix ${\bf{K}}_{{\bf{x}}_{g}} = \mathbb{E}\{{\bf{x}}_{g} {\bf{x}}\herm _{g}\} = \mathbb{E}\{{\bf{W}}_g{\bf{d}}_g {\bf{d}}\herm _g{\bf{W}}\herm _g\} = {\bf{W}}_g{\bf{I}}_{L_g}{\bf{W}}\herm _g$, was limited to $L_g$.
%
%
Here we propose a different approach, where we relax the rank limitation on ${\bf{K}}_{{\bf{x}}_{g}}$ and remove the linear per-stream decodability requirement in~\eqref{const: per-stream rate}. 
Therefore, in general, this relaxation would require encoding across spatial dimensions, and hence, non-linear receiver processing not considered in this paper. 
However, we can use the result as an upper-bound on the performance and compare it with the iterative solution proposed in Section~\ref{sec:iterative_solution}. 

For the upper-bound approach, we consider a generalized transmission vector $\hat{\Bx}_g$ for group $g$, with a generalized covariance matrix $\hat{\BK}_{\Bx_g}$. Then, the received signal model in~\eqref{eq:received signal_stream_specific} can be re-written as
\begin{equation}\label{eq: recieved signal_upper_bound}
   \hat{\By}_k = {\bf{H}}_k \hat{\Bx}_g + \sum_{\bar g \neq g} {\bf{H}}_k \hat{\Bx}_{\bar g} + {\bf{z}}_k \; , 
\end{equation}
and the achievable sum-rate of user $k \in \mathcal{K}_g$ during the transmission of $\hat{\Bx}_g$ is
\begin{equation}
    R_k = \log\left|{\bf{I}}+{\bf{Q}}_{g}^{-1}{\bf{H}}_k \hat{\BK}_{\Bx_{g}} {\bf{H}}_k \herm\right|,
\end{equation}
where ${\bf{Q}}_{g} := \sum_{\bar g \neq g}{\bf{H}}_k \hat{\BK}_{\Bx_{\bar g}}{\bf{H}}_k \herm + \sigma_k^2{\bf{I}}$.
Accordingly, the \ac{WMMF} problem would change to 
\begin{subequations}\label{eq:WMMF_upperbound_nonconvex}
\begin{align}
\hat{\SfS}_0 \ : &\max_{\hat{\BK}_{\Bx_{g}}} \min_{g \in [G], k \in \mathcal{K}_g} \alpha_{g}R_k\\
    \textrm{s.t.} \quad &\sum_{g \in [G]} \textrm{Trace}(\hat{\BK}_{\Bx_{g}}) \leq P_{T} \; .
\end{align}
\end{subequations}
The objective function in $\hat{\SfS}_0$ is not convex. However, one can show that it can be written as the difference of convex functions as
\begin{align}\label{eq:DCF}
    R_{k} = &\log\bigg|\sum_{\bar g \in [G]}{\bf{H}}_k \hat{\BK}_{\Bx_{\bar g}}{\bf{H}}_k \herm + \sigma^2_k{\bf{I}}\bigg|- \log\left|{\bf{Q}}_{g}\right|, \quad \forall k \; . 
\end{align}
Using~\eqref{eq:DCF} in~\eqref{eq:WMMF_upperbound_nonconvex} and following similar steps as in section~\ref{sec:problemformulation_streamspecific}, the problem $\hat{\SfS}_{0}$ can be iteratively approximated with the following convex problem
\begin{subequations}\label{eq:WMMF_upperbound_convex}
\begin{align}
\hat{\SfS}_{1} \ : &\max_{\hat{\BK}_{\Bx_{g}},R} R\\
\textrm{s.t.} \quad & \sum_{\bar g \neq g} \mathrm{Trace}\left(\bar{{\bf{Q}}}_{g}^{-1}{\bf{H}}\herm _k\left(\hat{\BK}_{\Bx_{\bar g}}-\overline{{\bf{K}}}_{\Bx_{\bar g}}\right){\bf{H}}_k\right) +\log\left|\bar{{\bf{Q}}}_{g}\right| \nonumber \\
& \ \ - \log\bigg|\sum_{g \in [G]}{\bf{H}}_k \hat{\BK}_{\Bx_{g}}{\bf{H}}\herm _k + \sigma^2_k{\bf{I}}\bigg| + \frac{R}{\alpha_g} \leq 0, \!\! \quad \forall (g,k) \; , \\
&\sum_{g \in [G]}\mathrm{Trace}(\hat{\BK}_{\Bx_{g}}) \leq P_{T}\label{const:power_upperbound} \; ,
\end{align}
\end{subequations}
where $\bar{{\bf{Q}}}_{g} := \sum_{\bar g \neq g}{\bf{H}}_k \overline{{\bf{K}}}_{\Bx_{\bar g}}{\bf{H}}\herm _k + \sigma_k^2{\bf{I}}$, and $\overline{{\bf{K}}}_{\Bx_{ g}}$ denotes the covariance matrix of the transmitted signal to group $g$ at the previous iteration. Since~\eqref{eq:WMMF_upperbound_convex} is convex, it can be directly handled by generic solvers such as CVX. The required procedure is quite similar to what we followed in Section~\ref{solution: CVX_streamspecific}, and is outlined in Algorithm~\ref{algorithm:MMF_cvx_upperbound}. Finding the solution to problem $\hat{\SfS}_1$ follows a similar approach to the \ac{SDR} method proposed in~\cite{MMF_QoS_MGMC_lefteris_2008}, and hence, its computation complexity is in the order of $\mathcal{O}((N_{\mathrm T} + \sum_{k \in [K]}N_k)^{6})$.

\begin{algorithm}
\SetAlgoLined
\KwResult{$\hat{\BK}_{\Bx_{g}}$, $R$}

 Choose random $\hat{\BK}_{\Bx_{g}}$ matrices such that~\eqref{const:power_upperbound} is met; 
 
 \While{ convergence is not met}{
 set $\overline{{\bf{K}}}_{\Bx_{g}} \gets \hat{\BK}_{\Bx_{g}}, \quad \forall g$,
 
 update $\hat{\BK}_{\Bx_{g}}, R$ using $\overline{{\bf{K}}}_{\Bx_{g}}$, from~\eqref{eq:WMMF_upperbound_convex}; 
 }
 \caption{CVX-based solution for upper-bound}
 \label{algorithm:MMF_cvx_upperbound}
\end{algorithm}

\vspace{-5pt}
\section{Simulation Results} \label{sec:simulation}

We use MATLAB simulations to compare the complexity and performance of the proposed beamforming solution with state-of-the-art.
%
We consider a downlink communication setup with equal-sized groups with uniform group priority (i.e., $\alpha_g = 1, \ \forall g$) and AWGN noise with unit variance ($N_0 = 1$). The users are assumed to have the same number of $N_{R}$ receive antennas, and the maximum number of transmitted streams for each group (i.e., $L_g$) is also considered to be equal to $N_R$. The step size $\beta$ is set to $10^{-2}$. 
We also consider the scheme proposed in~\cite{MGMC-ADMM-based-fast-algorithm-2017} as a baseline for the single-antenna receiver scenario (the per-antenna power constraint in~\cite{MGMC-ADMM-based-fast-algorithm-2017} is relaxed in our simulations). Note that in all figures, the numbers on the arrows show the actual simulation time averaged over all the realisations. SDPT3 solver is selected for CVX, and all the simulations are performed on the same hardware platform.

\begin{figure*}[t]
\minipage{0.32\textwidth}
  \includegraphics[ width=\linewidth]{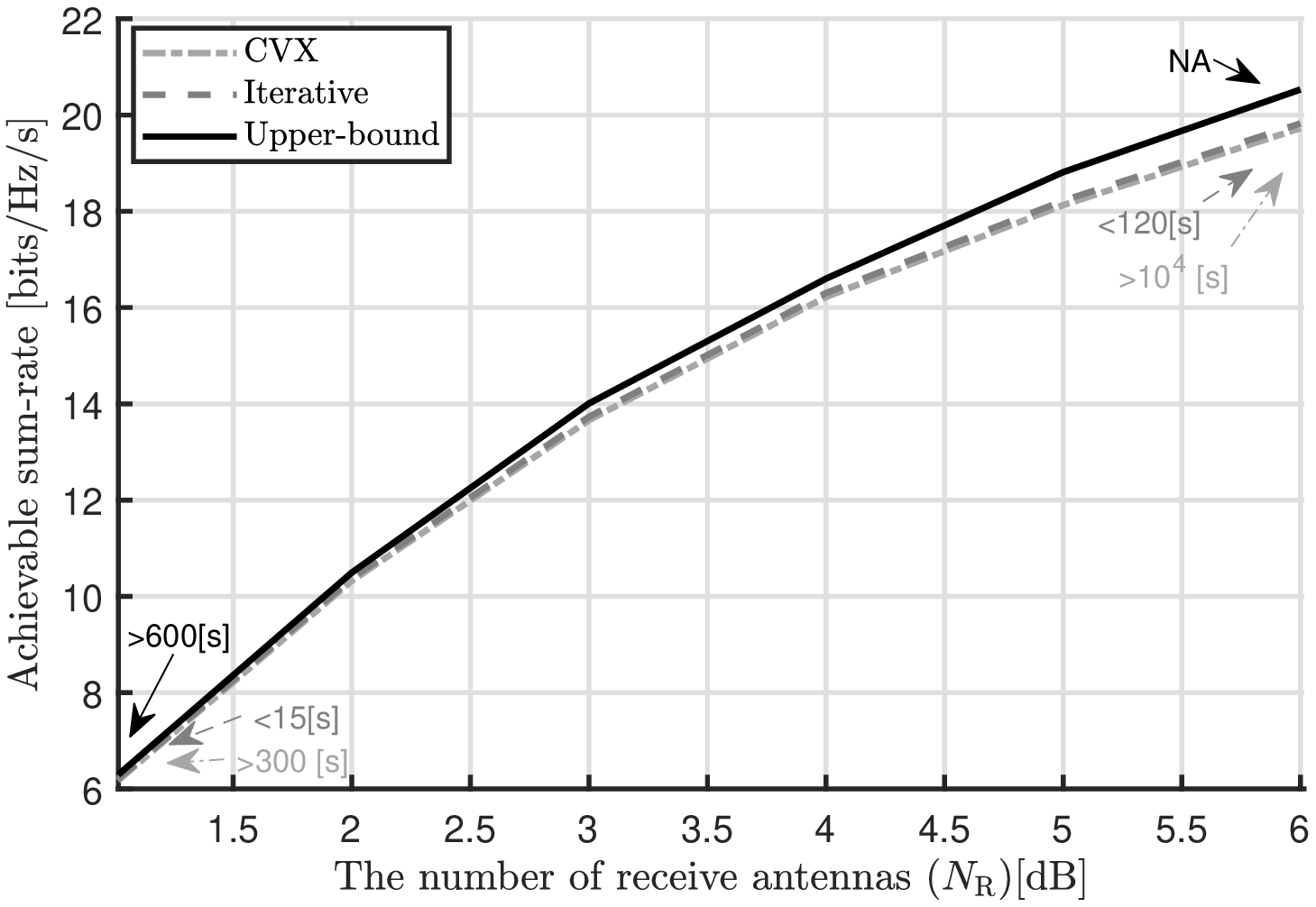}
  \caption{Achievable sum-rate vs. the number of receive antennas. $P_T = 10 \mathrm{dB}$, $G=3$, $N_{T} = 100$, $K = 15$}
\label{fig:simu-Nr}
\endminipage\hfill
\minipage{0.32\textwidth}
  \includegraphics[ width=\linewidth]{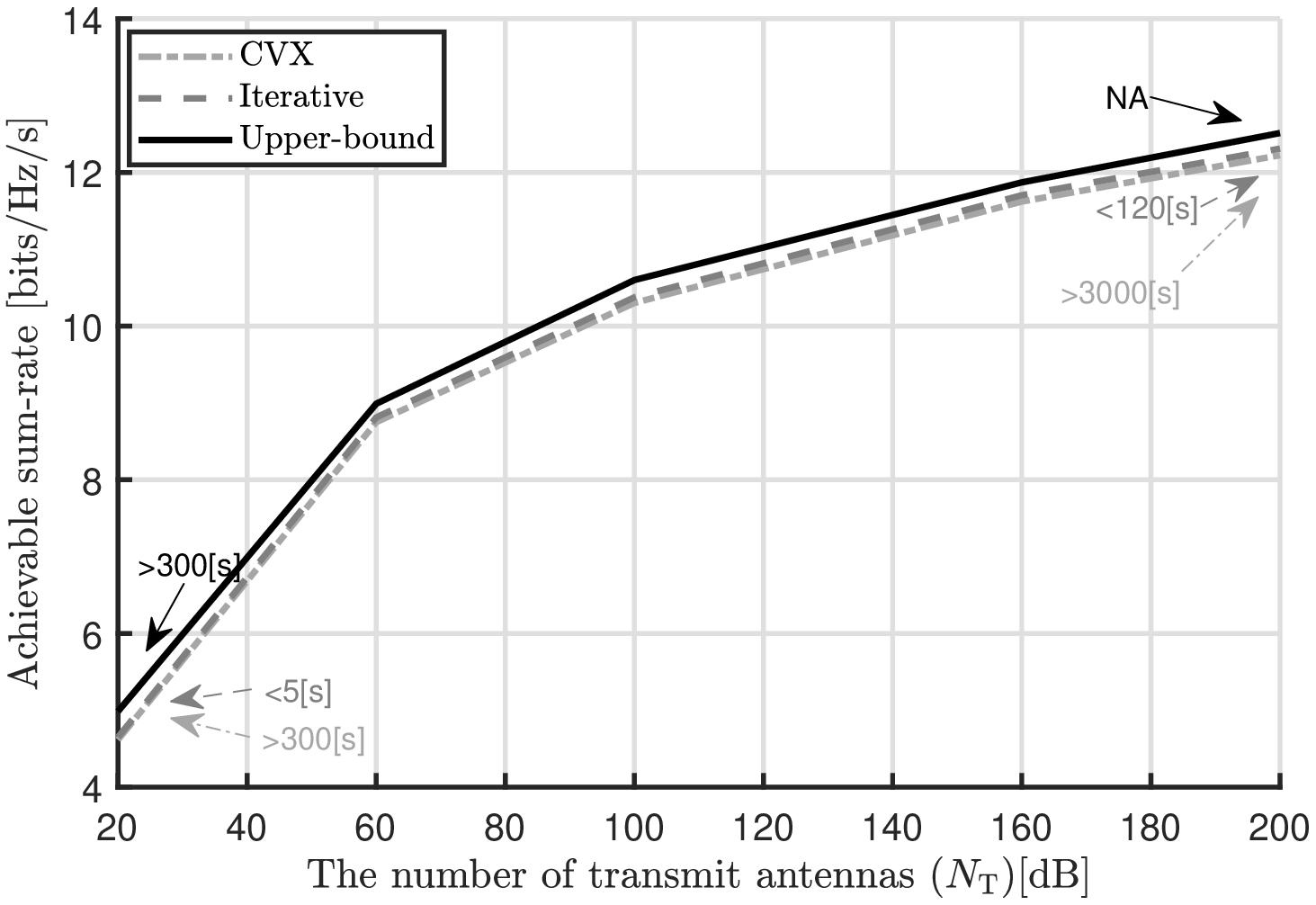}
  \caption{Achievable sum-rate vs. the number of transmit antennas. $P_T = 10 \mathrm{dB}$, $G=3$, $N_{R} = 2$, $K = 15$}
\label{fig:simu-Nt}
\endminipage\hfill
\minipage{0.32\textwidth}
  \includegraphics[ width=\linewidth]{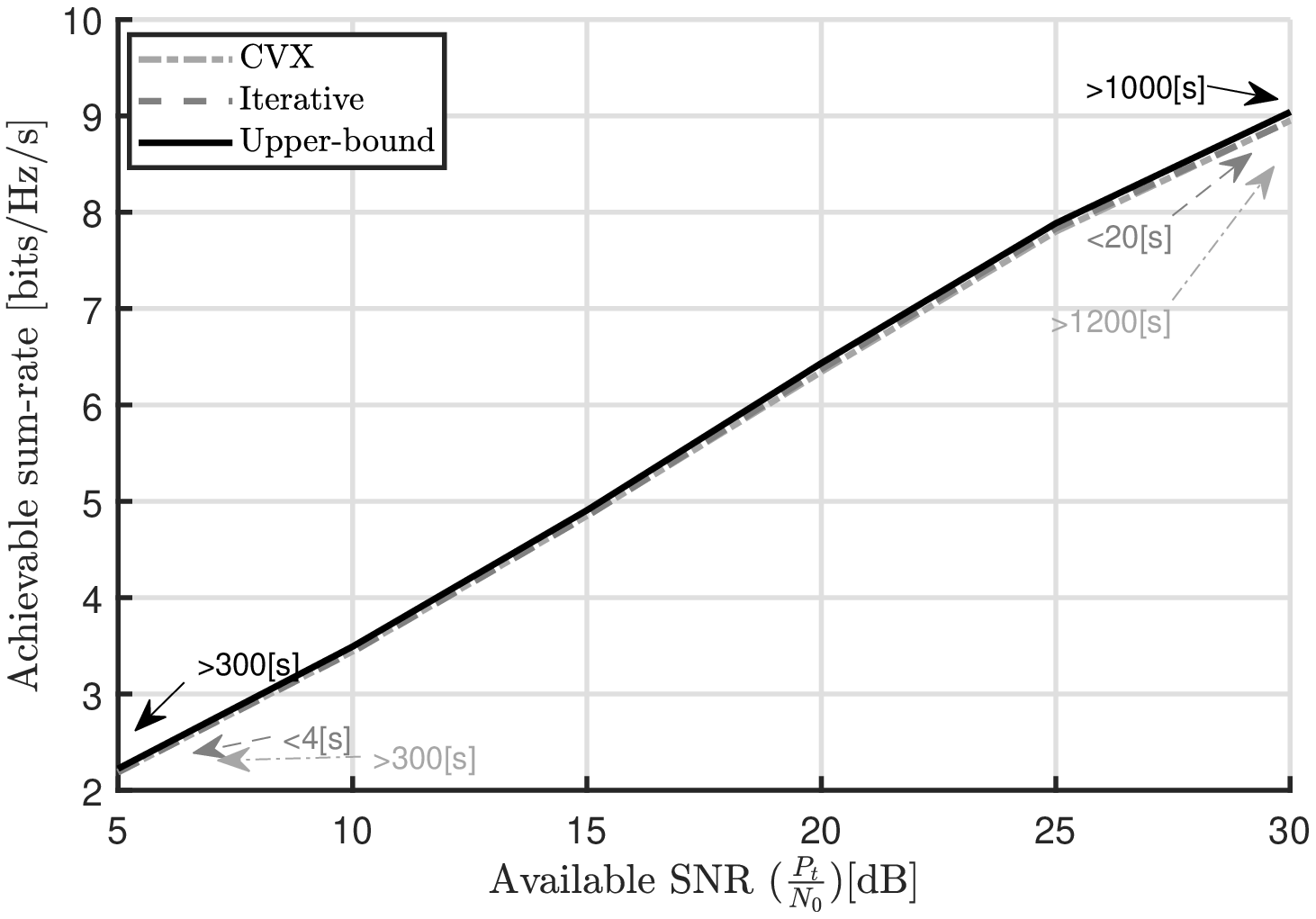}
  \caption{Achievable sum-rate vs. the available \ac{SNR}. $N_{T} = 10$, $G=3$, $N_{R} = 2$, $K = 15$}
\label{fig:simu-pt}
\endminipage
\end{figure*}

Fig.~\ref{fig:simu-Nr} compares the proposed iterative method with the CVX-based solutions, in terms of the achievable rate and the required convergence time, for different number of receive antennas at each user. As illustrated in the figure, both CVX-based and the proposed iterative solutions converge to the same optimal point but with considerably different convergence times. Moreover, it can be seen that the iterative solution performs fairly close to the upper-bound solution (which requires non-linear receiver implementation). However, the gap between the upper-bound and the linear solutions increases with the number of receive antennas $N_{R}$, as the rank of the transmit covariance matrix also increases with the same rate. It is also worth mentioning that due to the linearly growing complexity with respect to the number of receive antennas, the iterative solution is quite fast even for a large number of antennas, whereas CVX-based solutions are highly time-consuming.

Fig.~\ref{fig:simu-Nt} compares the performance of various methods for different number of transmit antennas. As can be seen, the proposed iterative method is significantly superior compared to the CVX-based solution, in terms of the required convergence time. Moreover, the gap between the linear solutions (CVX-based and iterative solutions) and the non-linear upper-bound method does not increase by the number of transmit antennas. In other words, the rank of the transmit covariance matrix is highly limited by the number of receive antennas (in the simulated scenario, this rank is almost always equal to $N_r$). It is also worth noting that the convergence time of the proposed method is more sensitive to the number of transmit antennas (compared to the number of receive antennas), confirming the previous discussions in Section~\ref{sec:iterative_solution}. 

Fig.~\ref{fig:simu-pt} depicts the achievable rate versus the available \ac{SNR} for the interference-limited scenario where the number of transmit antennas is not enough to support $N_{R} = 2$ streams for each group. Since the beamforming vectors are initialized such that $L_g=N_{\mathrm R}$ streams can be transmitted for each group, linear solutions need more iterations to converge to the optimal single-stream solution. This effect is more prominent in high-\ac{SNR} communications, as the interference terms are more dominant in this regime. Although the increased number of iterations results in an increase in the convergence time, still the iterative solution outperforms the CVX-based solution by a large margin.

\begin{figure}[t]
    \centering 
    \setlength\abovecaptionskip{-0.25\baselineskip}
    \includegraphics[width=1\columnwidth,keepaspectratio]{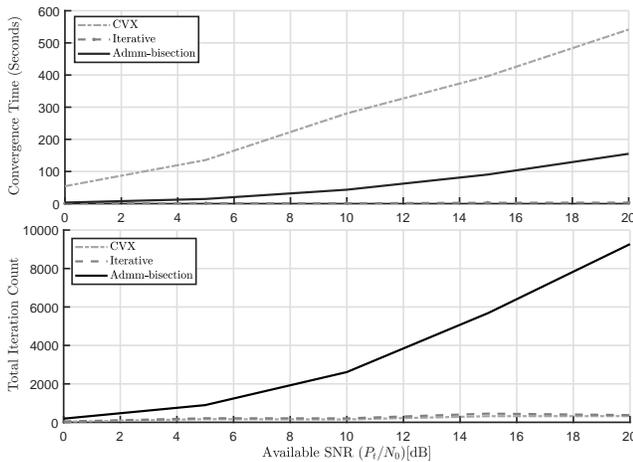}
    \caption{Convergence time and iteration count vs. the available \ac{SNR}. $G=3$, $K=15$, $N_{T}=20$, $N_{R}=1$.}
\label{fig:simu-converg}
\end{figure}

Finally, in Fig.~\ref{fig:simu-converg}, we have compared the convergence time and total iteration count (i.e., the total number of transmit and receive beamformer updates) for the proposed iterative method, the CVX-based solution, and the \ac{ADMM}-bisection method in~\cite{MGMC-ADMM-based-fast-algorithm-2017}. We have chosen the \ac{ADMM}-bisection as it has the best performance in our simulations (e.g., compared with~\cite{MMF_QoS_MGMC_lefteris_2008,SDR_gossian-randomization_MGMC_TSP_2014,MGMC-SCA-ganesh-2017,MGMCmmf}). As can be seen, even though the proposed solution is not originally designed for the single-antenna receiver scenario, its required iterations is smaller than the CVX-based solution and much smaller than the \ac{ADMM}-based solution (Note that the iteration counts of all the methods are increasing with the available \ac{SNR}. However, due to the different scaling, it is less visible for iterative and CVX-based solutions). This is because in our solution, the achievable rate is calculated in a closed-form, and hence, there is no need for bisection over the rate (as is the case for the \ac{ADMM}-based method as well as other works in the literature such as~\cite{MGMCmmf}). Similarly, the required convergence time of our iterative solution is smaller than the \ac{ADMM}-based method and much smaller than the CVX-based method. 



 \section{Conclusion and Future Work} \label{sec:conclusion}
In this paper, we proposed a low-complexity iterative method for the multi-stream multi-group multicasting problem with the weighted max-min fairness objective. In this method, the original non-convex and NP-hard problem is solved up to a locally optimal point by iterating between receive and transmit beamformer updates. 
Using \ac{KKT} conditions on the Lagrangian function, the optimal beamformer structure is derived as a function of dual variables. As some of the dual variables are highly-coupled and interdependent, an iterative sub-gradient method is used to find them efficiently. Finally, by finding the common achievable rate, the problem is solved directly without relying on bisection over the common rate.
Simulation results show that the proposed algorithm finds optimal beamformers much faster than generic solver-based methods. Potential extensions include non-perfect \ac{CSIT} and cell-free joint transmission over multiple transmitters.